\documentclass[twocolumn]{aastex7}
\usepackage{isotope}
\usepackage{soul}
\usepackage{hyperref}
\hypersetup{linkcolor=red,citecolor=cyan, filecolor=cyan,urlcolor=magenta}
\shortauthors{Psaltis et al.}
\graphicspath{{./}{figures/}}

\begin{document}

\title{Low-metallicity nova explosions: a site for weak \textit{rp}-process nucleosynthesis}

\author[0000-0003-2197-0797]{Athanasios Psaltis}
\affiliation{Department of Physics, Duke University, Durham, NC, 27710, USA}
\affiliation{Triangle Universities Nuclear Laboratory, Duke University, Durham, NC, 27710, USA}
\email[show]{psaltis.tha@duke.edu}
\author[0000-0002-9937-2685]{Jordi Jos\'e}
\affiliation{Departament de F\'isica, EEBE, Universitat Polit\`ecnica de Catalunya, c/Eduard Maristany 10, E-08930 Barcelona, Spain}
\affiliation{Institut d’Estudis Espacials de Catalunya, c/Esteve Terradas 1, E-08860 Castelldefels, Spain}
\email{jordi.jose@upc.edu}
\author[0000-0001-7731-580X]{Richard Longland}
\affiliation{Department of Physics, North Carolina State University,
Raleigh, NC, 27695, USA}
\affiliation{Triangle Universities Nuclear Laboratory, Duke University, Durham, NC, 27710, USA}
\email{rllongla@ncsu.edu}
\author[0000-0003-2381-0412]{Christian Iliadis}
\affiliation{Department of Physics \& Astronomy, University of North Carolina at Chapel Hill, NC 27599-3255, USA}
\affiliation{Triangle Universities Nuclear Laboratory, Duke University, Durham, NC, 27710, USA}
\email{iliadis@unc.edu}

\begin{abstract}

Classical novae are common cataclysmic events involving a binary system of a white dwarf and a main sequence or red giant companion star.
In metal-poor environments, these explosions produce ejecta different from their solar counterparts due to the accretion of sub-solar metallicity material onto the white dwarf.
In particular, it has been suggested that the nucleosynthesis flow in such low-metallicity nova explosions extends up to the Cu-Zn region, much beyond the expected endpoint, around Ca, predicted for solar-metallicity classical novae.
This behavior resembles a weak \textit{rp}-process, and such nuclear activity has never been observed in accreting white dwarf binaries with typical accretion flows.
In this work, we study the characteristics of the weak \textit{rp}-process for four nova models with metallicities $Z= 2\times 10^{-9}$, $10^{-7}$, $2\times 10^{-6}$, and $2\times 10^{-5}$, and explore the impact of the nuclear physics uncertainties via a Monte Carlo sensitivity study.
We identify nuclear reactions whose uncertainties affect the production of intermediate-mass nuclei under these conditions.
These reactions and relevant nuclear quantities are targets for measurements at stable or radioactive beam facilities to reduce their rate uncertainties.

\end{abstract}

\keywords{ Classical novae (251) --- Explosive nucleosynthesis (503) --- Nuclear Astrophysics (1129) --- Nuclear reaction cross sections (2087)}


\section{Introduction}
\label{sec:intro}

Classical novae~\citep[for some reviews]{Starrfield2008, Starrfield2012, Starrfield2016, Jose2008, Jose2016, Chomiuk2021} are stellar thermonuclear explosions, occurring at a rate of $\approx$ 30-80 per year in our Galaxy~\citep{Shafter2017}.
Only $\approx$ 5-10 are detected annually\footnote{Public database of Galactic novae ``galnovae'' available at \href{https://github.com/Bill-Gray/galnovae/}{https://github.com/Bill-Gray/galnovae/}}, as many are obscured by interstellar dust.
These events occur in binary systems, where a white dwarf accretes material from a main sequence or a red giant companion.
The material from the companion star accumulates onto the CO or ONe white dwarf and triggers a thermonuclear explosion on its surface.
The nucleosynthesis endpoint in classical novae is predicted by models and confirmed by observations to be around $A \sim 40$ (calcium).
Each nova outburst ejects only $\sim 10^{-7}-10^{-4}~M_\odot$ of material into the interstellar medium, which is why they are not considered major contributors to Galactic Chemical Evolution.
However, they can be significant galactic contributors of $\isotope[13][]{C}, \isotope[15][]{N},$ and $\isotope[17][]{O}$~\citep{Jose2016}.
In addition, $\isotope[7][]{Li}$, which originates from the decay of $\isotope[7][]{Be}$, has been suggested to be mainly produced in nova explosions~\citep{Starrfield2020}, but this remains a subject of ongoing debate~\citep{Jose2020}.

Early in Galactic history, low- and even zero-metallicity (Pop III) stars~\citep{Stacy2013, Stacy2016, Klessen2023} likely formed binary systems that could have produced a unique type of stellar explosion, with an energy output between a classical nova\footnote{In this paper, we use the following nomenclature for classical nova explosions with different metallicities: \textit{Solar-metallicity} nova refers to a white dwarf accreting solar-composition material. \textit{Primordial nova} describes a system where a Pop III star transfers material to a white dwarf. \textit{Low-metallicity} nova encompasses all the other cases where the companion star has a sub-solar metallicity.} and a supernova.
This low-metalicity binary system is expected to behave differently from a classical nova with solar metalicities.
\cite{Jose2007b} proposed that, unlike classical novae that produce elements only up to Ca, low-metallicity novae could synthesize heavier elements, reaching the Cu-Zn region through a sequence of $(p,\gamma)$ radiative captures and $\beta^+$ decays.
The distinguishing features between the two environments are the higher peak temperature reached in the latter, along with the dredging-up of freshly synthesized material from the interior of the red giant, which is transferred in the white dwarf envelope, triggering a breakout from the hot CNO cycles~\citep{Iliadis2015b, Wiescher2010}.

Beyond the work of~\cite{Jose2007b}, the impact of metallicity on nova explosions has been explored in several studies~\citep{Piersanti2000, Starrfield2000, Shen2007, Shen2009, Chen2019}, though typically not for values below $Z = 10^{-6}$.
Most recently,~\cite{Kemp2024} investigated the role of novae in Galactic Chemical Evolution using binary population synthesis models across a range of metallicities from $Z = 10^{-4}$ to $3 \times 10^{-2}$.
Their results suggest that both the nova rate and the amount of material ejected into the interstellar medium are inversely proportional to the metallicity of the system.

Low-metallicity environments—such as the Milky Way halo, the Magellanic Clouds, globular clusters, and dwarf galaxies—are expected to host a larger fraction of binary systems capable of producing nova explosions.
For example, the Large and Small Magellanic Clouds have mean metallicities of $\mathrm{[Fe/H] \approx -0.33(30)}$ and $\mathrm{[Fe/H] \approx -0.83(30)}$, respectively~\citep{Hocde2023}.
In such environments, novae are predicted to eject more massive shells, potentially enhancing their contribution to Galactic Chemical Evolution.
These low-metallicity novae may have also left an imprint in the inventory of presolar stardust grains~\citep{Amari2001}.

Extending the analysis of~\cite{Jose2007b}, which modeled novae with accreted material matching the lowest stellar metallicities observed, we find that a more detailed study is necessary.
Their reaction network, coupled to the hydrodynamic code, was limited to 270 nuclei and 1,400 reactions (see Section~\ref{sec:nucleo} for details).
Many thermonuclear reaction rates in the mass range $A= 30-50$ have large uncertainties that eventually affect the final abundance pattern.
Investigating these nuclear uncertainties in a hydrodynamical model is prohibitive due to the computational expense of the calculations.
Instead, it is preferred to perform post-processing calculations using an extended nuclear reaction network and a temperature and density profile extracted from a hydrodynamical simulation~\citep[see the work by][]{Iliadis2002}.

Although classical novae have been observed from $\gamma$-rays to radio waves~\citep{Chomiuk2021}, there is no direct detection of a Pop III star (Z=0)\footnote{Recent observations using JWST have pointed to possible signatures of Pop III stars at redshift $z= 10.6$~\citep{Maiolino2024}. One of the most metal-deficient stars observed, SMSS J031300.36-670839.3 has a [Fe/H]= -7.1~\citep{Keller2014}. $\mathrm{[Fe/H]}= \mathrm{\log_{10}\left(\frac{N_{Fe}}{N_H}\right)_\star- \log_{10}\left(\frac{N_{Fe}}{N_H}\right)_\odot}$.}, or a Pop III binary system.
\cite{Hartwig2015} has provided an estimate of $\sim 10^6$ Pop III survivors in the halo of the Milky Way, which have not yet been observed but would likely be the target of the next generation of space telescopes at high-$z$~\citep{Nakajima2022}.
To strengthen our theoretical framework, it is crucial to improve our understanding of the nucleosynthesis processes and observational signatures involving low-metallicity nova explosions.
By refining our theoretical predictions, we aim to establish a solid foundation for interpreting the observed abundances in these systems, ensuring that we are prepared to draw meaningful conclusions from future detections with next-generation space telescopes.

In the present work, we aim to answer the following two questions: (a) What does the nucleosynthesis in low-metallicity novae look like, and (b) How do the nuclear physics uncertainties affect it?
To achieve this, we have calculated the nucleosynthesis of four 1D low-metallicity novae models, with $Z= 2\times 10^{-9}, 10^{-7}$, $2\times 10^{-6}$, and $2\times 10^{-5}$, building on the work of~\cite{Jose2007b}.
These calculations use an extended nuclear network coupled to a state-of-the-art reaction rate library.
We have also performed Monte Carlo reaction network studies to explore the impact of thermonuclear reaction rate uncertainties on the resulting nucleosynthesis.
Answering these questions will help us better understand nucleosynthesis early in Galactic history and offer a guide for the experimental nuclear physics community to pursue measurements of key reactions of astrophysical interest at stable and radioactive ion beam facilities.

This paper is structured as follows: in Section~\ref{sec:astro} we discuss the specifics of low-metallicity nova explosions and present the different hydrodynamical profiles that we selected for our study. Section~\ref{sec:nucleo} will provide a detailed description of the nucleosynthesis processes that occur in these nova explosions, and in Section~\ref{sec:mc}, we present our reaction network setup for the Monte Carlo sensitivity study. In Sections~\ref{sec:results} and~\ref{sec:conclusions} we present our results, and finally, conclude and discuss our results.

\section{Hydrodynamical simulations of low-metallicity novae explosions}
\label{sec:astro}

\begin{deluxetable*}{ccccc}[ht!]
\tablecaption{Model parameters for the nova simulations in this study. All nova models involve an ONe white dwarf of $\mathrm{M^{ini}_{WD}= 1.35~M_{\odot}}$, $\mathrm{R_{WD}= 2260~km}$, $\mathrm{L_{ini}= 10^{-2}~L_{\odot}}$ and $\mathrm{M_{acc} = 2 \times 10^{-10}~M_{\odot}~yr^{-1}}$. \label{tab:1}}
\tablehead{ \colhead{Property} & \colhead{Model Z2e-9} & \colhead{Model Z1e-7} & \colhead{Model Z2e-6 } & \colhead{Model Z2e-5} }
\startdata
Composition of accreted material& Solar/$10^7$ & Solar/$2 \times 10^5$ & Solar/$10^4$ & Solar/$10^3$   \\
 Metallicity of accreted material & $Z=2 \times 10^{-9}$ & $Z=10^{-7}$ & $Z=2\times 10^{-6}$ & $Z=2\times 10^{-5}$\\
$\mathrm{M_{env}~(10^{-5}~M_{\odot}})$ & 3.36 & 1.78 & 1.38 & 1.28  \\
$\mathrm{P_{max}~(10^{19}~dyn~cm^{-2})}$ & 33.9 & 18.1 & 14.1 & 13.2 \\
$\mathrm{T_{max}~(MK)}$ & 466 & 385 & 364 & 359  \\
$\mathrm{M_{ejec}~(10^{-5}~M_{\odot})}$ & 2.71 & 1.43 & 1.11 & 1.03 \\
$\mathrm{v_{ejec}~(km~s^{-1})}$ & 4239 & 3950 & 4068 & 4140
\enddata
\end{deluxetable*}

In the present work, we revisit the two low-metallicity nova models presented in~\cite{Jose2007b}, and we expand our reach in the metallicity phase space by including two more models.
Their metallicity values $Z$, are $2 \times 10^{-9}$, $10^{-7}$, $2\times 10^{-6}$, and $2\times 10^{-5}$. For more details, see Table~\ref{tab:1}. These correspond to spectroscopic values [Fe/H]= -7, -5.4, -4, and -3, respectively.
The lowest metallicity nova model was selected to match one of the most metal-deficient star observed, SMSS
J031300.36-67083~\citep{Keller2014}.
Models Z1e-7 and Z2e-6 correspond to Models A and B of~\cite{Jose2007b}.
All models were computed using the one-dimensional implicit Lagrangian hydrodynamic code \textsc{Shiva}~\citep[see][for details]{Jose1998, Jose2016}, which has been widely used to model classical nova explosions and Type I X-ray bursts.
The code self-consistently couples hydrodynamics with an extensive nuclear reaction network to simulate thermonuclear runaways and the resulting nucleosynthesis.
\textsc{Shiva} incorporates energy generation from nuclear reactions, convective transport, and the effects of a degenerate equation of state, making it a robust tool for studying nova outbursts across different metallicities.

As discussed in~\cite{Jose2007b}, the choice of a rather massive white dwarf ($\mathrm{M_{WD}= 1.35~M_\odot}$) stems from the need to explore the nucleosynthetic endpoint of these explosions, since more massive white dwarfs lead to higher peak temperatures during the explosion phase, due to the higher pressure at the base of the envelope, and eventually a more violent outburst.
For the present work, we did not compute models for a variety of WD masses since our main goal was to study the extent of nuclear activity in low-metallicity novae.
A comprehensive characterization of the outcomes of novae as a function of white dwarf mass, initial luminosity, and mass-accretion rate is beyond the scope of this paper and will be addressed in future work.

Table~\ref{tab:1} shows the parameters of the low-metallicity novae models computed in this work.
The quantities $\mathrm{M_{wd}}$ and $\mathrm{R_{wd}}$ are the initial values of the white dwarf mass and radius, after relaxation of the initial model, just when accretion sets in; $\mathrm{L_{ini}}$ is the initial white dwarf luminosity;
$\mathrm{M_{acc}}$ is the mass accretion rate, and $\mathrm{M_{env}}$ is the final mass of the accreted envelope when the thermonuclear runaway begins. The quantities $\mathrm{P_{max}}$ and $\mathrm{T_{max}}$ are the maximum pressure and temperature achieved at the ignition shell;
$\mathrm{M_{ejec}}$ and $\mathrm{v_{ejec}}$ are the mass and mean-velocity of the ejecta.
The models reach peak temperature $\mathrm{T_{max}}$,  between 359 and 466~MK, which is significantly higher than the typical range of 100–300~MK observed in solar-metallicity novae~\citep{Jose2016}.

\begin{figure*}[ht!]
    \centering
    \includegraphics[width=\textwidth]{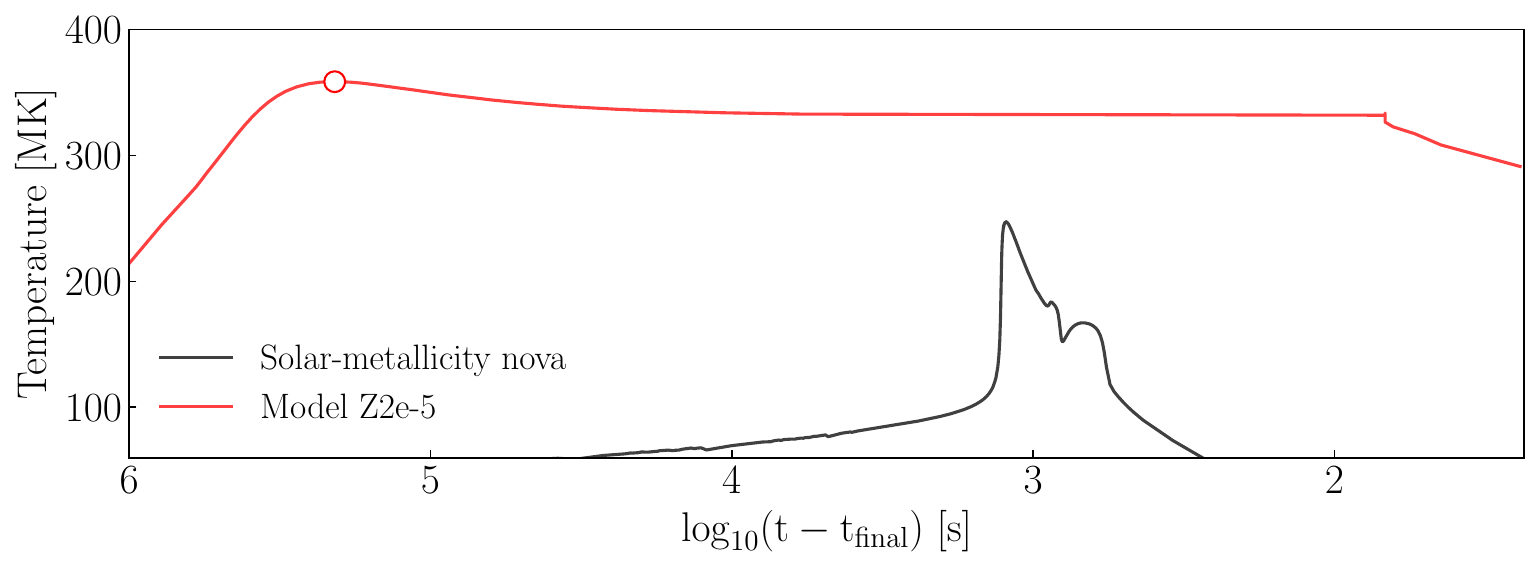}
    \caption{ Temperature evolution of the innermost shell for a low-metallicity nova (Model Z2e-5, red) and a solar-metallicity
    ONe nova (black).
    Note the significantly higher peak temperature and longer duration of high temperatures in the low-metallicity
    model.
    The white circle indicates the beginning of the post-processing calculation at $\mathrm{T=T_{max}}$.}
\label{fig:profile}
\end{figure*}

From the nuclear physics point of view, the key difference in the models considered here to the usually published solar-metallicity models is not only that the former achieve higher peak temperatures, but the high temperatures are also maintained for a much longer time.
We illustrate that feature in Figure~\ref{fig:profile}, where the time evolution of Model Z2e-5 is compared with a typical solar-metallicity ONe novae also computed using the \textsc{Shiva} code~\citep{Jose2016,Jose1998}.

Although we assume that the donor star is unevolved in all models, such that the accreted material reflects the original low-metallicity composition, we note that if the donor were evolved, the composition of the transferred material could be enriched.
The hydrodynamic models computed assume that the stream of material accreted by the white dwarf has the same metallicity as the secondary star (see Table 1 for the specific values adopted in each model).
However, inspired by multidimensional models of mixing at the core-envelope interface through Kelvin-Helmholtz hydrodynamic instabilities, some material is assumed to be dredged up into the innermost layer of the envelope with a
characteristic timescale given by the convective turnover time, $\tau_{conv} \sim 10$~s, as soon as the envelope becomes fully convective~\citep{Glasner2012,Glasner2007,Glasner1997, Casanova2018, Casanova2016, Casanova2011a, Casanova2011b, Casanova2010}.
Tests performed with different choices of convective turnover time, from 10~s—very similar to the values reported in the multidimensional calculations of Glasner \textit{et al.} and Casanova \textit{et al.}—to 100~s, have shown little impact on the results.

Injecting $\isotope[12][]{C}$ into the burning region triggers the thermonuclear runaway via the $\isotope[12][]{C}(p,\gamma)\isotope[13][]{N}$ reaction, making dredge-up of white dwarf and accreted material a crucial component to achieve high nuclear reaction activity.
Without the injection of $\isotope[12][]{C}$ into the burning region, even a hot explosion would not lead to a CNO breakout if the initial composition had very low metallicity.
In Table~\ref{tab:2} we present the white dwarf core composition, which is based on ~\cite{Ritossa1996}.
During this dredge-up phase, the mass fraction $\mathrm{X_i}$ of each isotope is approximated using the following:
\begin{equation}
    \mathrm{X_i = \frac{M_1 X_{old} + M_2 X_{dredge}}{M_1 + M_2}}
\end{equation}
where $\mathrm{X_{old}}$ is the mass fraction at the previous timestep, $\mathrm{X_{dredge}}$ is the composition of the outermost white dwarf layers, as listed in Table~\ref{tab:2}, $M_2 = M_1 \frac{\Delta t}{\tau}$, $M_1$, is the mass of the innermost envelope shell. $\tau = \tau_{conv} = 10$~s and $\Delta t$ is the integration timestep. This mixing prescription is assumed both in the hydrodynamical models~\citep{Jose2007b} and during the nucleosynthesis post-processing, with modifications that account for the difference between integrating a single-zone thermodynamic profile and a full 1D model, which we discuss in the next section.

\begin{deluxetable}{cc}[!ht]
\tablewidth{10pt}
\tablecaption{Chemical composition at the outermost layers of an ONe white dwarf based on~\cite{Ritossa1996}.\label{tab:2}}
\tablehead{ \colhead{Isotope} & \colhead{Mass Fraction, $X_i$} }
\startdata
$\isotope[12][]{C}$& 9.16e-3   \\
$\isotope[16][]{O}$& 5.11e-1   \\
$\isotope[20][]{Ne}$& 3.13e-1   \\
$\isotope[21][]{Ne}$& 5.98e-3   \\
$\isotope[22][]{Ne}$& 4.31e-3   \\
$\isotope[23][]{Na}$& 6.44e-2   \\
$\isotope[24][]{Mg}$& 5.48e-2   \\
$\isotope[25][]{Mg}$& 1.58e-2   \\
$\isotope[26][]{Mg}$& 9.89e-3   \\
$\isotope[27][]{Al}$& 1.08e-2
\enddata
\tablecomments{The composition is taken at mass
$\mathrm{M=1.17~M_{\odot}}$.}
\end{deluxetable}

\section{Nucleosynthesis}
\label{sec:nucleo}

The nucleosynthesis in low-metallicity novae explosions was first discussed in~\cite{Jose2007b}.
In that work, a network of 270 nuclei ($\isotope[1][]{H}$ to $\isotope[75][]{As}$) was followed through a network of 1,400 nuclear reactions, using reaction rates based on experimental information for many stable-target a few radioactive-target reactions, such as for $\isotope[21][]{Na}(p,\gamma)\isotope[22][]{Mg}$~\citep{Bishop2003}.
In this work, we extend the reaction network to 457 nuclei (n to $\isotope[88][]{Kr}$) and 4,969 nuclear processes.
All thermonuclear reaction rates were taken from the STARLIB reaction rate library~\citep{Sallaska2013} (version 6.10)\footnote{The STARLIB thermonuclear reaction rate library can be found at \href{https://starlib.github.io/Rate-Library/}{https://starlib.github.io/Rate-Library/}.}.
For the reactions that do not have any experimental information, a theoretical rate is used, based on the statistical model calculations of the \textsc{Talys} code~\citep{Goriely2008}.

Each network calculation was computed for the duration of the time evolution of the $(T -\rho)$ profile extracted from the hydrodynamic simulation, between roughly 26~min for model Z2e-9 to 58~h for model Z2e-5.
Note that the starting point $t=0$ of each calculation is when the model reaches $\mathrm{T_{max}}$.
At the end of each calculation, we decay all species with $t_{1/2} < 1~h$ to their stable counterparts.
The network was integrated using a semi-implicit, second order Runge-Kutta method (often referred to as ``Wagoner's method'')~\citep{Wagoner1969, Longland2014}, ensuring consistency with the hydrodynamical models.
However, we took a slightly different approach with the dredge-up scheme.
In the hydrodynamical models, mixing begins when the envelope becomes fully convective and continues until the end of the simulation.
However, in our post-processing approach, we terminated the mixing process much earlier.
Specifically, we applied dredging for the following durations: Z2e-9 7.5~s (1574~s), Z1e-7 40~s (1476~s), Z2e-6 75~s (1656~s), and Z2e-5 75~s (556~s).
The numbers in parentheses represent the total duration of dredging in the hydrodynamical simulation.
Prolonging the injection of white dwarf core material for the same duration as in the hydrodynamical models—ranging from thousands of seconds to a few days—would have caused the final abundances to be dominated by that material since there is no intra-shell mixing in our post-processing framework.
To address this, we tuned the dredge-up time so that the final mass fraction pattern closely matches the mean composition of the ejecta in the hydrodynamical simulation.
We emphasize that this tuning is only used to reproduce the overall bulk enrichment, as our one-zone post-processing framework uses the temperature and density evolution from the \textsc{Shiva} code for only the innermost shell.
The primary goal of our study is to quantify the impact of nuclear reaction rate uncertainties on the production of intermediate-mass nuclei, rather than to make direct comparisons with observed values.
In this context, we are confident that this adjustment does not affect the overall conclusions of our study.

\begin{figure*}[hbpt!]
    \centering
    \includegraphics[width=.75\textwidth]{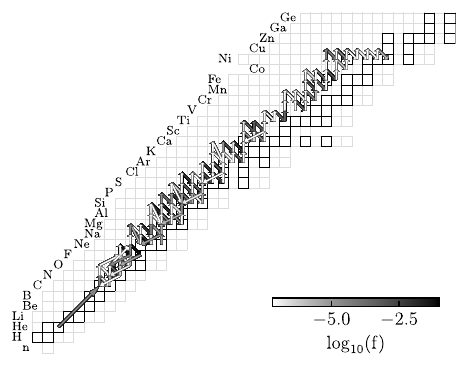}
    \caption{Time-integrated reaction flux for Model Z1e-7. The black and grey boxes correspond to the stable and radioactive isotopes of the network, respectively. The color of each flux arrow indicates its magnitude, with darker arrows showing stonger fluxes. See the text for details.}
\label{fig:flowZ1e-7}
\end{figure*}

Figure~\ref{fig:flowZ1e-7} presents the total time-integrated reaction flux for Model Z1e-7.
The net flux from isotope $i$ to isotope $j$ is defined as $f_{ij}= \int \dot{X}_{i \rightarrow j} - \dot{X}_{j \rightarrow i}~dt$, where $\dot{X}_{i \rightarrow j}$ is the rate of change of the mass fraction of isotope $i$ by all the reactions that convert it into isotope $j$.
The strongest flux include the $(p,\gamma)$ reactions followed by $\beta^+$ decays.

In contrast to the \textit{rp}-process in Type I X-ray bursts~\citep{Schatz1998}, after the breakout of the hot CNO cycle, in the low-metallicity nova scenario we do not see a sequence of $(\alpha,p)$ and $(p,\gamma)$ reactions, known as the $\alpha p$-process~\citep{Wallace1981}.
According to Table~\ref{tab:1}, at the peak temperatures achieved in the models, 359-466 MK, the Coulomb barrier penetrability for $\alpha$-particles remains very low, preventing $\alpha$-captures from becoming dominant.
Instead, we get a $(p, \gamma) - \beta^+$ pattern, close to the valley of stability.
This also indicates that the critical reactions identified in this study can be accessible by experimental studies.

Nucleosynthesis in a low-metallicity nova setting resembles a ``\textit{weak rp-process}'', whose extent is between a classical nova and a full \textit{rp}-process, which has a nucleosynthesis endpoint at the SnSbTe mass region~\citep{Schatz1998}.
Similar nuclear activities have been reported in other studies, such as in~\cite{vanWormer1994}, and~\cite{Fisker2008}, but in the context of accreting neutron stars.
Such nuclear activity has not been reported for accreting white dwarfs with typical accretion rates of $\mathrm{M_{acc} = 2 \times 10^{-10}~M_{\odot}~yr^{-1}}$.
\cite{Glassner2009} have reported breakout from the hot CNO cycle for very low accretion rates of
$\mathrm{M_{acc} = 10^{-11}~M_{\odot}~yr^{-1}}$, which, however, are not supported by observations.
The ability of low-metallicity nova models to develop a weak \textit{rp}-process depends on two key factors: the high peak temperature ($\mathrm{T_{peak} > 3 \times 10^8}$~K) and, more importantly, the mixing between the low-metallicity envelope and the outer layers of the white dwarf.
The low metallicity of the accreted material results in reduced nuclear activity during the accretion phase, prolonging its duration and allowing more mass to accumulate.
This leads to higher pressures at the base of the envelope.
A key factor in this context is the time required for the envelope to become fully convective, as this marks the onset of dredge-up.
Once convection sets in, fresh $\isotope[12][]{C}$ is mixed into the envelope, creating thermodynamic conditions favorable for a powerful explosion and enabling the envelope to reach the high peak temperatures required for ignition.
The explosion is then triggered by the $\isotope[12][]{C}(p, \gamma)\isotope[13][]{N}$ reaction, initiating the thermonuclear runaway.

For Model Z1e-7 the reaction flow cannot proceed beyond $A=67$ and $\isotope[67][]{Ge}$ $\beta^+$ decays back to $\isotope[67][]{Zn}$, while for Models Z2e-6 and Z2e-5 the endpoint is slightly lower in mass number.
The lowest metallicity nova (Model Z2e-9) exhibits a weak \textit{rp}-process with the shortest nucleosynthesis endpoint around cobalt -- maximum element where $\mathrm{\log_{10}(X_i/X_\odot) > 0}$.
This is due to its helium-rich composition (initially X(H)=0.314 and X(He)=0.371; see Table~\ref{tab:abundances} for the full initial mass fractions. These should not be confused with the composition of the outermost layers of the ONe white dwarf, which are provided in Table~\ref{tab:2}.), and the temperature is not high enough for $\alpha$-capture reactions to move the reaction flow to heavier masses.
This behavior is likely a numerical artifact of the one-zone approach, which lacks the convective mixing present in hydrodynamic models, leading to insufficient amount of hydrogen fuel.
In realistic conditions, convection replenishes hydrogen in the envelope by mixing material from cooler regions.
Hydrodynamic simulations of low-metallicity novae consistently show that protons are not consumed entirely, enabling continued nucleosynthesis.

\section{The impact of thermonuclear reaction rate uncertainties}
\label{sec:mc}

Monte Carlo sensitivity studies are widely used in the literature for a variety of nucleosynthesis scenarios, \textit{e.g.} Big Bang Nucleosynthesis (BBN)~\citep{Iliadis2020}, X-ray bursts~\citep{Parikh2008}, $\nu p$-process~\citep{Nishimura2019}, \textit{s}-process~\citep{Cescutti2018}, \textit{r}-process~\citep{Mumpower2016}, weak \textit{r}-process~\citep{Psaltis2022}, \textit{i}-process~\citep{Denissenkov2021} and the $\gamma$-process~\citep{Rauscher2016}.
Monte Carlo studies offer several advantages over traditional methods for assessing the impact of nuclear physics uncertainties in astrophysical environments, such as individually varying reaction rates (\textit{e.g.},~\cite{Iliadis2002}).
They allow a full propagation of nuclear physics uncertainties to final abundances, which can be summarized with statistically meaningful metrics.
They also naturally include the possible impact of multiple reactions impacting the nucleosynthesis of an isotope: while one rate might significantly alter an abundance, a second rate could either compensate for or amplify this change.

We evolve our reaction network, detailed in Section~\ref{sec:nucleo}, using the thermodynamic conditions outlined in Section~\ref{sec:astro}, initiating our calculations at the peak temperature ($\mathrm{T_{max}}$, Table~\ref{tab:1}) to preserve hydrogen fuel until the onset of the thermonuclear runaway.
This approach serves as a tool for performing one-zone post-processing calculations while ensuring consistency with the results obtained using the hydrodynamic code \textsc{Shiva}.

Each model underwent 10,000 post-processing nucleosynthesis calculations, varying reaction rates based on the uncertainties listed in the STARLIB rate library~\citep{Sallaska2013}.
The reaction rates in STARLIB have lognormal probability densities at each temperature $T$ and can be described by:
\begin{equation}
f(r) = \frac{1}{\sigma \sqrt{2\pi}} \frac{1}{r} e^{-(\ln r -\mu)^2/(2\sigma^2)},~\mathrm{for~0<r<\infty}
\end{equation}
where $r$ is the reaction rate. The recommended (median), low, and high rates are given by $r_{median}= e^\mu, r_{low}=e^{\mu-\sigma}$, and $r_{high}= e^{\mu+\sigma}$, respectively~\citep{Longland2010}. The factor uncertainty $f.u.$, which determines the 68\% coverage probability, is given by $f.u.= e^\sigma$~\citep{Iliadis2015}.

In each MC calculation $i$, we assign a random variation factor $p_{ij}$ to each nuclear interaction.
This factor is drawn from a normal (Gaussian) distribution with a mean of $\mu=0$ and a variance $\sigma^2=1$, \textit{i.e.} $p_{ij} \sim N(0,1)$, where $N(0,1)$ denotes the standard normal distribution.
This factor is then used to calculate the sampled rate of reaction $j$ using the following:
\begin{equation}\label{eq:2}
    r_{ij} =  r_{median, j} [f.u]_j^{p_{ij}}
\end{equation}

For forward and reverse reactions, we apply the same variation exponent $p$.
Each sampled reaction rate is multiplied by a factor $[f.u]_j^{p_{ij}}$ at each temperature (T = 0.01-10~GK), maintaining the temperature dependence of the rate uncertainty.
For a more detailed discussion regarding the sampling procedure in a Monte Carlo nucleosynthesis study, we refer the interested reader to~\cite{Longland2012}.

\begin{figure}[hbpt!]
    \centering
    \includegraphics[width=0.5\textwidth]{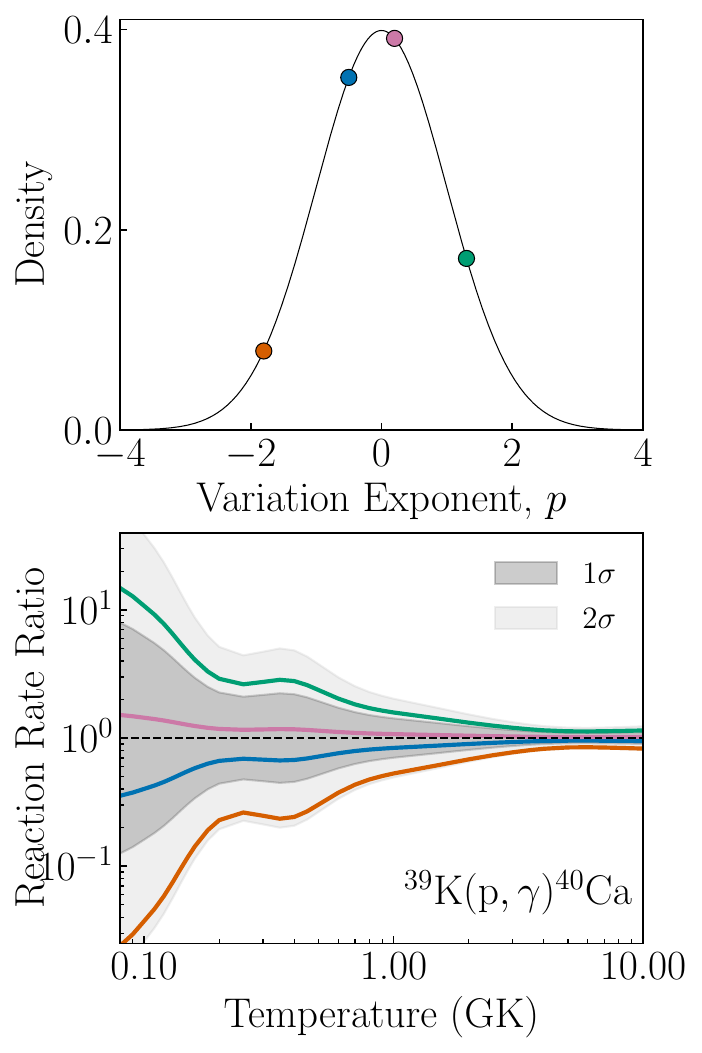}
    \caption{ (Top) Four draws from the Normal distribution for the exponent factor $p$. (Bottom) The resulting reaction rates compared to the median for the $\isotope[39][]{K}(p, \gamma)\isotope[40][]{Ca}$ reaction rate. See the text for details.}
\label{fig:mc-drawing}
\end{figure}

To visually represent the above procedure, we show an example for the $\isotope[39][]{K}(p, \gamma)\isotope[40][]{Ca}$ reaction rate in Figure~\ref{fig:mc-drawing}.
In the top panel, four draws of the rate variation factor $p$, which is normally distributed, are illustrated.
In the bottom panel, these points are translated into reaction rates that are compared with the median using Equation~\ref{eq:2}.
For example, the blue point in the top panel of Figure~\ref{fig:mc-drawing} represents a sample with $p = -0.5$.
Next, the sampled reaction rate is calculated using Equation~\ref{eq:2}, which yields the blue line in the bottom panel of Figure~\ref{fig:mc-drawing}.
The difference between the sampled and the median rate is not constant; it varies with the temperature-dependent uncertainty.
Similarly to the blue point, the orange, violet, and green points correspond to draws with $p= -1.8, 0.2,$ and 1.3, respectively.

\begin{figure*}[hbpt!]
    \centering
    \includegraphics[width=\textwidth]{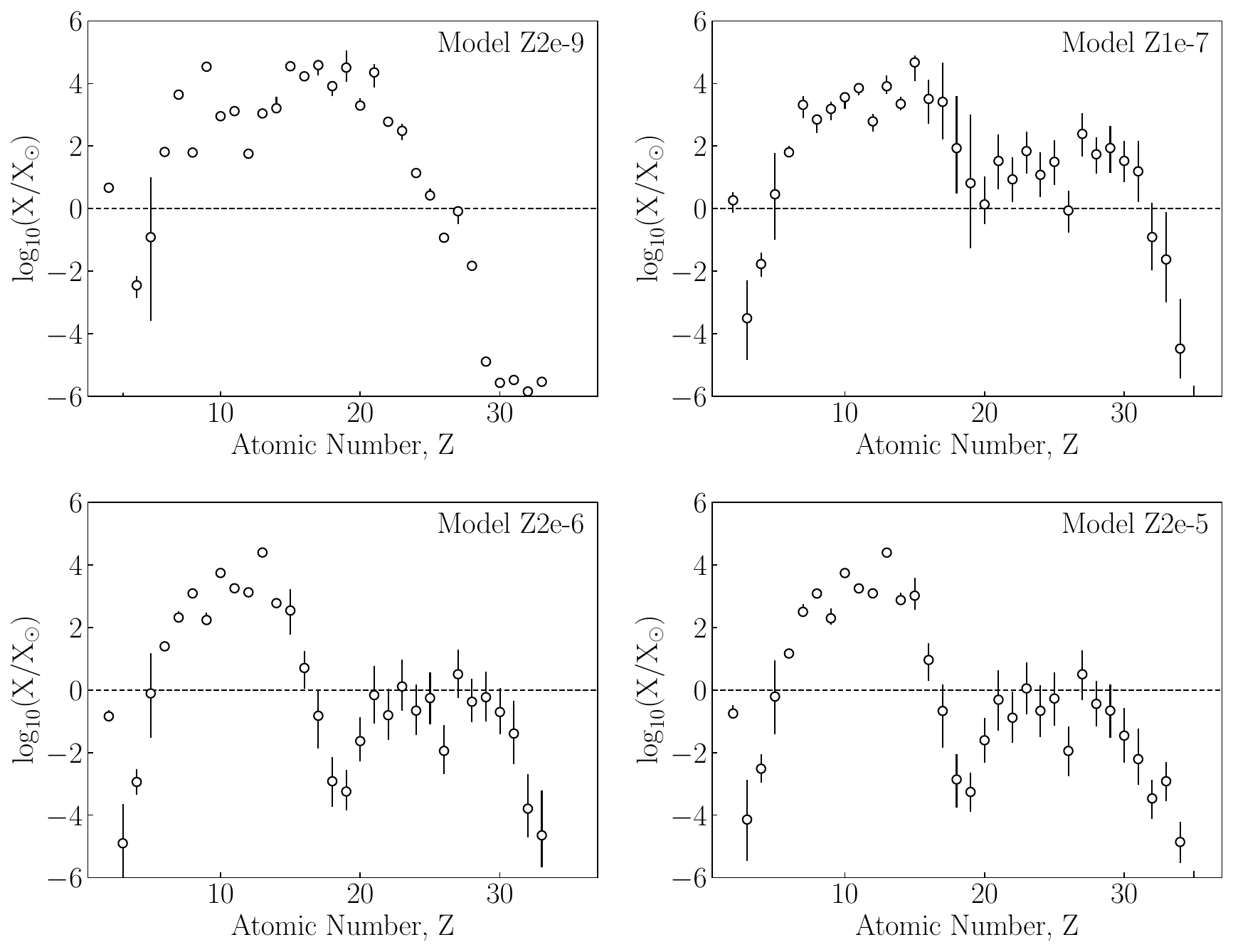}
    \caption{Overproduction plots for each model as a ratio of the logarithm of the abundance of each element to its solar value~\citep{Lodders2020} -- $\mathrm{\log_{10}(X/X_{\odot})}$. The 1$\sigma$ uncertainties (68\% coverage probability) are due to the variation of thermonuclear reaction rates using a Monte Carlo sampling. See Section~\ref{sec:mc} for a detailed discussion.}
\label{fig:overproduction-MC}
\end{figure*}

In the Monte Carlo nucleosynthesis framework, each reaction network calculation results in an abundance pattern, and their aggregate can help construct a probability density for each element or isotope.
Figure~\ref{fig:overproduction-MC} shows the final overproduction factors, $\mathrm{\log_{10}(X/X_\odot)}$, for the stable isotopes for all models with their $1\sigma$ uncertainty (68 \% coverage probability) from the variation of the thermonuclear reaction rates.

\begin{figure*}[ht!]
    \centering
    \includegraphics[width=\textwidth]{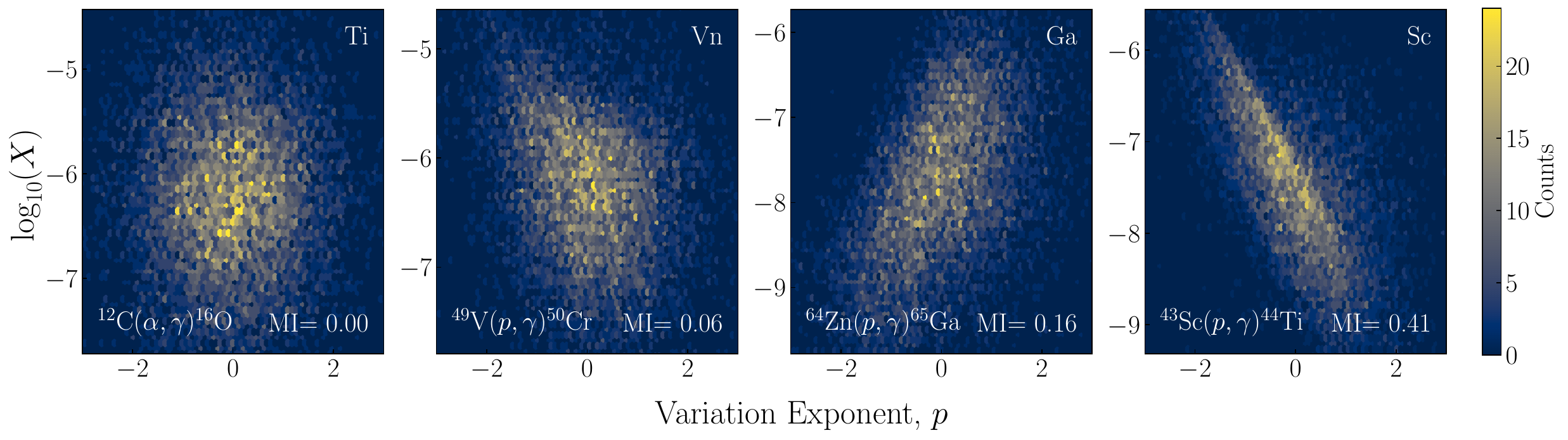}
    \caption{Correlations between elemental mass fractions and reaction rates for four representative cases in Model Z1e-7.
    Each panel highlights a specific element (indicated in the top right corner: titanium, vanadium, gallium, and scandium) and the corresponding reaction with its mutual information (MI) value (displayed at the bottom).
    The panels are arranged in order of increasing MI value from left to right, illustrating the strength of the correlation.
    A reaction is considered ``important'' if its MI exceeds the threshold of 0.10, as no significant correlations are visually discernible for reactions with MI below this value.}
\label{fig:mi}
\end{figure*}

The impact of a thermonuclear reaction rate uncertainty is found by determining the correlation between the variation exponent $p$ and the final mass fraction of an element $X_k$.
In the present work, we use the mutual information (MI) metric, as it was first introduced for MC nucleosynthesis in~\cite{Iliadis2020}.
It has the advantage, compared to commonly used metrics in Monte Carlo studies, such as the Pearson (linear)~\citep{Nishimura2019} and Spearman (monotonic)~\citep{Psaltis2022} correlation coefficients, to capture correlations that are \textit{neither linear nor monotonic}.

For two random variables, in our analysis the simulated Monte Carlo mass fractions $\{X_{1k}, X_{2k}, X_{3k}, \ldots\}$ and the sampled rates of the j\textsuperscript{th} reaction,  which are drawn based on its uncertainty, $\{r_{1j}, r_{2j}, r_{3j}, \ldots \}$ the mutual information can be written as\footnote{We use the \texttt{mutual\_info\_regression} module from the \texttt{scikit-learn}~\citep{scikit-learn} package to calculate the MI values.}:
\begin{equation}
    \mathrm{MI = \sum_X \sum_r P(X,r) \log\left[ \frac{P(X,r)}{P(X)P(r)}\right]}
\end{equation}
where $\mathrm{P(X)}$ and $\mathrm{P(r)}$ are the marginal distributions of the mass fraction and the reaction rate, and $\mathrm{P(X,r)}$ is the joint probability density.

For the results in Section~\ref{sec:results}, we use a critical limit of MI = 0.09 to define an ``important'' reaction, one whose uncertainty needs to be reduced.
Since MI has no upper bound, we visually inspected the plots to identify meaningful changes in correlation.
Figure~\ref{fig:mi} shows four example cases from Model Z1e-7 (see Table~\ref{tab:reactions} for a complete list), arranged according to the magnitude of the MI metric.
The leftmost panel shows the variation of the $\isotope[12][]{C}(\alpha,\gamma)\isotope[16][]{O}$ reaction, which shows no visible correlation with the mass fraction of Ti.
The second panel displays a small correlation with MI of less than 0.09.
The last two examples correspond to reactions identified as ``important'' for the model, where a correlation can be clearly seen.
It can be observed that below MI = 0.09, no correlation is discernible visually.
Therefore, in this work, we use this cutoff to define an important reaction, and only these reactions are presented in the tables.
\section{Results}
\label{sec:results}

This section highlights the key findings from our Monte Carlo sensitivity analysis.

\subsection{Reactions that affect elemental abundances}

We present in Table~\ref{tab:reactions} the reactions whose mutual information (MI) metric is above 0.09 for elements with a mean mass fraction $\mathrm{X_i > 10^{-10}}$ and with abundance uncertainties of more than a factor of 3.

As expected, most identified reactions are proton captures, $(p,\gamma)$.
The $\isotope[15][]{O}(\alpha,\gamma)\isotope[19][]{Ne}$ reaction, which is a breakout from the hot CNO cycles affects the production of Co and Ni in multiple models, because it controls how fast the material flows from the CNO region to heavier mass isotopes.
The $\isotope[18][]{F}(p,\alpha)\isotope[15][]{O}$ closes the hot CNO cycle and affects the production of fluorine in models Z2e-6 and Z2e-5.
The $(p,\gamma)$ reactions on the following target nuclei affect multiple elements in many models: $\isotope[30][]{P}$, $\isotope[33,34][]{S}$,
$\isotope[43][]{Sc}$, $\isotope[44, 46][]{Ti}$, $\isotope[50][]{Cr}$, $\isotope[53][]{Mn}$, $\isotope[54][]{Fe}$, $\isotope[63][]{Cu}$, and $\isotope[64,65][]{Zn}$.
For most of the aforementioned isotopes, the $(p,\gamma)$ reaction rates are based on theoretical estimates using the statistical Hauser-Feshbach model, and no experimental data are available.
In the STARLIB library, such reactions have a factor uncertainty ($f.u.$) of 10 for the whole temperature grid.

\startlongtable
\begin{deluxetable*}{cccccc}
\tablecaption{Reactions with $\mathrm{MI\geq 0.09}$ that affect isotopic mass fractions more than a factor of 3 in any model.
The reactions indicated with a $\diamond$ include a radioactive target nucleus. \label{tab:reactions}}
\tablehead{\colhead{Reaction} & \colhead{Affected Element} & \colhead{Model Z2e-9 MI} & \colhead{Model Z1e-7 MI} & \colhead{Model Z2e-6 MI} & \colhead{Model Z2e-5 MI}}
\startdata
$\isotope[11][]{B}(\alpha,n)\isotope[14][]{N}$ & B &  & 0.09 & 0.10 &  \\
$\isotope[11][]{C}(\alpha,p)\isotope[14][]{N}$$\diamond$ & B & 0.20 &  &  &  \\
$\isotope[11][]{C}(\alpha,n)\isotope[14][]{O}$$\diamond$ & B & 0.11 & 0.34 & 0.40 & 0.57 \\
$\isotope[15][]{O}(\alpha,\gamma)\isotope[19][]{Ne}$$\diamond$ & Fe &  & 0.09 &  & 0.10 \\
 & Ni &  & 0.19 & 0.11 & 0.11 \\
 & Co &  & 0.11 &  & 0.09 \\
 & Ca &  &  &  & 0.10 \\
$\isotope[17][]{F}(p,\gamma)\isotope[18][]{Ne}$$\diamond$ & N &  & 0.16 &  &  \\
 & O &  & 0.13 &  &  \\
$\isotope[18][]{F}(p,\alpha)\isotope[15][]{O}$$\diamond$ & F &  &  &  & 0.16 \\
$\isotope[30][]{P}(p,\gamma)\isotope[31][]{S}$$\diamond$ & Si & 1.19 &  &  &  \\
 & Cl & 0.20 &  & 0.19 & 0.17 \\
 & S &  & 0.25 & 0.78 & 0.64 \\
 & P &  & 0.54 & 0.76 & 0.23 \\
 & Ar &  &  & 0.14 & 0.14 \\
$\isotope[33][]{S}(p,\gamma)\isotope[34][]{Cl}$ & Cl & 0.14 &  & 0.14 & 0.11 \\
 & Ar &  &  & 0.09 & 0.09 \\
$\isotope[34][]{S}(p,\gamma)\isotope[35][]{Cl}$ & Cl & 0.12 &  & 0.18 & 0.15 \\
 & Ar &  &  & 0.14 & 0.11 \\
$\isotope[37][]{Ar}(p,\gamma)\isotope[38][]{K}$$\diamond$ & K & 0.15 &  &  &  \\
$\isotope[38][]{K}(p,\gamma)\isotope[39][]{Ca}$$\diamond$ & K & 0.54 & 0.09 &  &  \\
$\isotope[39][]{K}(p,\gamma)\isotope[40][]{Ca}$ & K & 0.13 &  & 0.16 & 0.13 \\
$\isotope[42][]{Ca}(p,\gamma)\isotope[43][]{Sc}$ & Sc & 0.20 &  &  &  \\
$\isotope[43][]{Ca}(p,n)\isotope[43][]{Sc}$ & Sc & 0.13 &  &  &  \\
$\isotope[43][]{Sc}(p,\alpha)\isotope[40][]{Ca}$$\diamond$ & Ca &  &  & 0.08 & 0.09 \\
$\isotope[43][]{Sc}(p,\gamma)\isotope[44][]{Ti}$$\diamond$ & Sc & 0.20 & 0.41 & 0.20 & 0.35 \\
 & Ca &  &  & 0.09 &  \\
$\isotope[44][]{Ti}(p,\gamma)\isotope[45][]{V}$$\diamond$ & Ti &  & 0.16 & 0.11 & 0.11 \\
$\isotope[45][]{Ti}(p,\gamma)\isotope[46][]{V}$$\diamond$ & Sc &  & 0.13 &  &  \\
$\isotope[46][]{Ti}(p,\gamma)\isotope[47][]{V}$ & V & 0.31 &  &  &  \\
 & Ti &  & 0.12 & 0.09 & 0.09 \\
$\isotope[47][]{Ti}(p,n)\isotope[47][]{V}$ & V & 0.28 &  &  &  \\
$\isotope[47][]{V}(p,\gamma)\isotope[48][]{Cr}$$\diamond$ & V & 0.09 &  &  &  \\
$\isotope[48][]{V}(p,n)\isotope[48][]{Cr}$$\diamond$ & V & 0.09 &  &  &  \\
$\isotope[49][]{V}(p,\gamma)\isotope[50][]{Cr}$$\diamond$ & V &  &  &  & 0.10 \\
$\isotope[48][]{Cr}(p,\gamma)\isotope[49][]{Mn}$$\diamond$ & Cr &  & 0.11 &  &  \\
$\isotope[50][]{Cr}(p,\gamma)\isotope[51][]{Mn}$ & Cr &  & 0.15 & 0.12 & 0.12 \\
$\isotope[52][]{Mn}(p,\gamma)\isotope[53][]{Fe}$ & Mn &  & 0.11 &  & 0.09 \\
$\isotope[53][]{Mn}(p,\gamma)\isotope[54][]{Fe}$$\diamond$ & Mn &  & 0.16 & 0.13 & 0.14 \\
$\isotope[54][]{Fe}(p,\gamma)\isotope[55][]{Co}$ & Fe &  & 0.18 & 0.12 & 0.12 \\
$\isotope[55][]{Co}(p,\gamma)\isotope[56][]{Ni}$$\diamond$ & Co &  & 0.18 &  &  \\
$\isotope[63][]{Cu}(p,\gamma)\isotope[64][]{Zn}$ & Cu &  & 0.09 & 0.13 &  \\
 & Zn &  & 0.11 & 0.16 & 0.13 \\
$\isotope[64][]{Zn}(p,\gamma)\isotope[65][]{Ga}$ & Ga &  & 0.16 & 0.12 &  \\
$\isotope[65][]{Zn}(p,\gamma)\isotope[66][]{Ga}$$\diamond$ & Ga &  & 0.16 & 0.15 &  \\
$\isotope[70][]{Ge}(p,\gamma)\isotope[71][]{As}$ & As &  & 0.09 &  &  \\
\enddata
\end{deluxetable*}

\subsection{Signatures of low-metallicity nova explosions}

The imprint of low-metallicity nova explosions is an exciting yet underexplored topic in observational astrophysics.
In this context, we will briefly discuss three potential ways to detect signatures of such explosions: through the study of Globular Clusters (GCs), extragalactic and galactic halo novae explosions, and presolar stardust grains.

High-resolution spectroscopy offers a powerful tool for investigating novae in low-metallicity environments. UV observations can provide elemental yields, while near-IR spectroscopy offers limited isotopic information.
Recently, \cite{Evans2024} reported the first infrared (IR) observation of a recurrent nova in the LMC, identifying silicon emission lines.

Globular Clusters are ideal environments to study stellar populations and their nucleosynthesis~\citep{Ashman2008, Iliadis2016}, and they can provide a viable environment to detect low-metallicity novae, given that they host an older population of stars.
Unfortunately, there have been only two confirmed nova detections in Galactic GCs; in M80 (T Sco 1660)~\citep{Sawyer1938} and in NGC 6402 (M14)~\citep{Sawyer1964}.
Given the small statistics, the inferred frequencies for novae in GCs vary widely from $\mathrm{5 \times 10^{-4}~yr^{-1}~GC^{-1}}$~\citep{Doyle2019} to $\mathrm{0.05~yr^{-1}~GC^{-1}}$~\citep{Henze2013}.

In addition to the Magellanic Clouds, the Milky Way halo and its surrounding dwarf galaxies are also expected to host low-metallicity binaries.
Our results, particularly those in Figure~\ref{fig:overproduction-MC}, suggest that low-metallicity nova explosions can overproduce a wide range of elements.
However, significant nuclear physics uncertainties (Section~\ref{sec:mc}) currently limit the precision with which we can constrain their contributions.
High-resolution spectroscopic observations hold promise for identifying the chemical imprints of these explosions.
In cases where isotopic abundances can be extracted through high-resolution IR spectroscopy, even tighter constraints on this nucleosynthesis scenario could be achieved.

Regarding presolar stardust grains~\citep{Amari2001, Zinner2014}, \cite{Jose2007b} proposed that models Z1e-7 (A) and Z2e-6 (B) could explain some unusual isotopic ratios found in Type-X grains.
The models we computed in this work yield very low $\isotope[14][]{N}/\isotope[15][]{N}$ ratios, even smaller than those observed in SiC-X grains, while reproducing the $\isotope[26][]{Al}^g/\isotope[27][]{Al}$ and $\isotope[30][]{Si}/\isotope[28][]{Si}$ ratios within uncertainties, based on the variation of reaction rates.
For titanium isotopes, all models predict substantial production of $\isotope[44][]{Ti}$ and $\isotope[46][]{Ti}$ relative to stable $\isotope[49][]{Ti}$.
Our results agree with~\cite{Jose2007b}, however, we cannot draw any definite conclusion due to the significant nuclear physics uncertainties.
Presolar grains are expected to survive in the interstellar medium for roughly 1~Gy before being incorporated into the presolar cloud about 4.8~Gy ago.
This implies that any grain originating from a low-metallicity nova would have had to condense no more than 5.8~Gy ago.
While the Milky Way has likely experienced at least two infall events~\citep{Spitoni2021}, the likelihood of a nova explosion in a low-metallicity binary significantly contributing to the pollution of the protosolar nebula remains low.
Nonetheless, given these uncertainties, such an event is not entirely out of the question.

\section{Conclusions and Discussion}
\label{sec:conclusions}

Low-metallicity novae are intriguing astrophysical environments that can produce elements up to Cu-Zn in metal-poor environments (Galactic halo, LMC, and SMC, dwarf galaxies, and GCs), including the early history of the Milky Way.

In the framework of the present work, we explored the nucleosynthesis that occurs in that scenario, using four models based on 1D hydrodynamic models calculated with the code \textsc{Shiva}.
Their nucleosynthesis flow is surpassing the standard, classical novae and resembles a weak \textit{rp}-process, which has only been reported in accreting neutron stars.
We also studied the impact of thermonuclear reaction rate uncertainties using a Monte Carlo approach, identifying several reactions that need to be further investigated experimentally.
While more advanced multi-dimensional methods are now feasible~\citep{Casanova2016, Jose2020}, they fall outside the scope of this study.
We show that, by adding a simple mixing scheme, 1D models can be used to capture the impact of nuclear physics uncertainties on this nucleosynthesis scenario.

Although low-metallicity nova explosions have yet to be observed, we anticipate that, in light of the findings presented in this work, a more systematic effort to target novae in globular clusters, the Galactic Halo, and dwarf galaxies—using current and future space-based telescopes—will provide valuable high-resolution spectroscopic data.
These observations will offer a unique opportunity to compare and refine our nucleosynthesis models.

The goal of the nuclear physics community should be to reduce the uncertainties of the reactions highlighted in Table~\ref{tab:reactions}, with upcoming experiments at stable and radioactive ion beam facilities to illuminate the low-metallicity novae contribution to the Galactic matter inventory.

\begin{acknowledgements}
The TUNL authors acknowledge support from U.S. Department of Energy, Office of Science, Office of Nuclear Physics, under Award Number DE-SC0017799 and Contract Nos. DE-FG02-97ER41033 (TUNL), DE-FG02-97ER41041 (UNC), and DE-FG02-97ER41042 (NCSU).
JJ acknowledges support from
the Spanish MINECO grant PID2023-148661NB-I00, the E.U. FEDER funds, and the AGAUR/Generalitat
de Catalunya grant SGR-386/2021.
The authors would like to thank S. Shore and N. Evans for fruitful discussions.
\end{acknowledgements}

\software{\texttt{IPython}~\citep{ipython}, \texttt{Jupyter}~\citep{jupyter}, \texttt{matplotlib}~\citep{matplotlib}, \texttt{numpy}~\citep{numpy}, \texttt{scikit-learn}~\citep{scikit-learn}, scientific colour maps~\citep{crameri}}

\appendix

\startlongtable
\begin{deluxetable}{ccccc}
\tablecaption{Initial elemental mass fractions for the different low-metallicity nova models extracted from the hydrodynamic code \textsc{Shiva} at $\mathrm{T=T_{max}}$. This is the starting point of the post-processing nucleosynthesis. For the composition of the outer layers of the white dwarf, see Table~\ref{tab:2}.}
\tablehead{ \colhead{Element} & \colhead{Model Z2e-9} & \colhead{Model Z1e-7} & \colhead{Model Z2e-6} & \colhead{Model Z2e-5}}
\startdata
H & 3.77e-01 & 6.06e-01 & 6.23e-01 & 6.65e-01 \\
He & 3.96e-01 & 3.93e-01 & 3.77e-01 & 3.34e-01 \\
Li & 2.93e-27 & 4.06e-27 & 3.70e-27 & 4.60e-27 \\
Be & 9.11e-14 & 3.83e-12 & 7.37e-12 & 3.31e-12 \\
B & 5.84e-19 & 3.25e-17 & 7.41e-17 & 6.97e-17 \\
C & 2.18e-07 & 5.39e-09 & 5.02e-09 & 8.58e-10 \\
N & 8.52e-07 & 3.59e-08 & 3.42e-08 & 5.68e-09 \\
O & 6.60e-02 & 4.21e-04 & 2.54e-04 & 2.28e-04 \\
F & 8.80e-03 & 9.71e-09 & 5.05e-09 & 3.95e-09 \\
Ne & 3.31e-02 & 3.20e-08 & 8.86e-09 & 4.85e-08 \\
Mg & 3.61e-02 & 2.59e-10 & 5.93e-11 & 1.01e-09 \\
Al & 1.82e-04 & 3.07e-11 & 8.78e-12 & 5.00e-11 \\
Si & 2.74e-03 & 3.90e-11 & 2.46e-11 & 9.81e-11 \\
P & 1.07e-02 & 1.29e-10 & 1.76e-09 & 3.30e-10 \\
S & 3.01e-02 & 6.58e-10 & 1.62e-10 & 1.84e-09 \\
Cl & 2.07e-02 & 2.41e-10 & 5.07e-11 & 9.73e-10 \\
Ar & 5.80e-03 & 2.11e-10 & 4.82e-11 & 6.62e-10 \\
K & 6.55e-03 & 1.11e-09 & 3.28e-10 & 1.96e-09 \\
Ca & 6.69e-03 & 5.38e-07 & 2.49e-07 & 4.52e-07 \\
Sc & 1.17e-04 & 3.51e-09 & 1.05e-09 & 4.02e-09 \\
Ti & 9.98e-05 & 1.61e-08 & 5.88e-09 & 1.58e-08 \\
V & 3.56e-05 & 1.61e-08 & 6.38e-09 & 1.01e-08 \\
Cr & 1.44e-05 & 5.88e-07 & 1.86e-07 & 2.90e-07 \\
Mn & 2.26e-06 & 9.64e-06 & 1.76e-06 & 5.63e-07 \\
Fe & 6.23e-06 & 1.25e-05 & 2.80e-06 & 5.44e-06 \\
Co & 1.55e-07 & 1.01e-05 & 3.28e-06 & 2.16e-06 \\
Ni & 4.24e-08 & 5.00e-05 & 1.01e-05 & 1.38e-05 \\
Cu & 2.90e-11 & 1.48e-06 & 9.36e-08 & 3.16e-07 \\
Zn & 2.25e-13 & 6.02e-07 & 1.77e-08 & 2.68e-08 \\
Ga & 8.99e-15 & 2.93e-09 & 5.99e-11 & 3.79e-10 \\
Ge & 1.02e-14 & 2.92e-11 & 1.60e-11 & 1.47e-10 \\
As & 1.05e-15 & 1.34e-13 & 1.86e-12 & 1.80e-11 \\
Se & 9.69e-17 & 8.57e-15 & 1.33e-13 & 1.28e-12 \\
Br & 1.78e-24 & 5.52e-21 & 8.94e-21 & 5.82e-19 \\
Kr & 1.13e-28 & 5.26e-26 & 4.57e-26 & 3.19e-23 \\
Rb & 2.19e-33 & 2.19e-33 & 2.19e-33 & 2.19e-33 \\
Sr & 7.40e-34 & 7.40e-34 & 7.40e-34 & 7.40e-34 \\
\enddata
\label{tab:abundances}
\end{deluxetable}

\startlongtable
\begin{deluxetable}{lcccc}
\tablecaption{Factor uncertainties (84\textsuperscript{th}/50\textsuperscript{th} percentile ratio) of elemental mass fraction for the studied models. We report factor uncertainties only for mean mass fractions $>10^{-10}$.\label{tab:variation}}
\tablehead{\colhead{} & \colhead{Model Z2e-9} & \colhead{Model Z1e-7} & \colhead{Model Z2e-6} & \colhead{Model Z2e-5}}
\startdata
Li & \nodata & \nodata & \nodata & \nodata\\
Be & \nodata & \nodata & \nodata & \nodata\\
B & 83.4 & 21.3 & 19.4 & 14.1\\
C & 1.2 & 1.6 & 1.1 & 1.1\\
N & 1.0 & 1.9 & 1.6 & 1.8\\
O & 1.3 & 1.5 & 1.0 & 1.0\\
F & 1.1 & 1.7 & 1.7 & 2.1\\
Ne & 1.2 & 1.3 & 1.0 & 1.0\\
Na & 1.2 & 1.4 & 1.2 & 1.3\\
Mg & 1.2 & 1.7 & 1.1 & 1.1\\
Al & 1.2 & 2.3 & 1.0 & 1.0\\
Si & 2.3 & 1.7 & 1.4 & 1.7\\
P & 1.3 & 1.6 & 4.8 & 3.7\\
S & 1.1 & 4.1 & 3.4 & 3.5\\
Cl & 1.5 & 17.9 & 6.6 & 7.1\\
Ar & 1.4 & 46.3 & 5.9 & 6.6\\
K & 3.5 & 156.1 & 4.9 & 4.1\\
Ca & 1.8 & 7.9 & 5.9 & 5.1\\
Sc & 1.8 & 7.2 & 8.3 & 8.6\\
Ti & 1.4 & 5.0 & 7.0 & 6.8\\
V & 1.6 & 4.2 & 7.1 & 6.7\\
Cr & 1.1 & 5.3 & 6.9 & 6.4\\
Mn & 1.6 & 4.8 & 6.5 & 6.8\\
Fe & 1.0 & 4.3 & 6.6 & 6.0\\
Ni & 1.0 & 3.5 & 5.5 & 5.3\\
Co & 1.1 & 4.7 & 6.2 & 5.9\\
Cu & \nodata & 5.2 & 6.4 & 6.8\\
Zn & \nodata & 4.4 & 5.7 & 7.8\\
Ga & \nodata & 9.2 & 10.9 & \nodata\\
Ge & \nodata & 12.7 & \nodata & \nodata\\
As & \nodata & 33.0 & \nodata & \nodata\\
\enddata
\label{tab:variations}
\end{deluxetable}

\bibliography{bibliography}{}

\begin{thebibliography}{}
\expandafter\ifx\csname natexlab\endcsname\relax\def\natexlab#1{#1}\fi
\providecommand{\url}[1]{\href{#1}{#1}}
\providecommand{\dodoi}[1]{doi:~\href{http://doi.org/#1}{\nolinkurl{#1}}}
\providecommand{\doeprint}[1]{\href{http://ascl.net/#1}{\nolinkurl{http://ascl.net/#1}}}
\providecommand{\doarXiv}[1]{\href{https://arxiv.org/abs/#1}{\nolinkurl{https://arxiv.org/abs/#1}}}

\bibitem[{S. {Amari} {et~al.}(2001){Amari}, {Gao}, {Nittler}, {Zinner}, {Jos{\'e}}, {Hernanz}, \& {Lewis}}]{Amari2001}
{Amari}, S., {Gao}, X., {Nittler}, L.~R., {et~al.} 2001, \bibinfo{title}{{Presolar Grains from Novae},} \apj, 551, 1065, \dodoi{10.1086/320235}

\bibitem[{K.~M. {Ashman} \& S.~E. {Zepf}(2008){Ashman} \& {Zepf}}]{Ashman2008}
{Ashman}, K.~M., \& {Zepf}, S.~E. 2008, {Globular Cluster Systems} (Cambridge University Press)

\bibitem[{S. Bishop {et~al.}(2003)Bishop, Azuma, Buchmann, Chen, Chatterjee, D'Auria, Engel, Gigliotti, Greife, Hernanz, Hunter, Hussein, Hutcheon, Jewett, Jos\'e, King, Kubono, Laird, Lamey, Lewis, Liu, Michimasa, Olin, Ottewell, Parker, Rogers, Strieder, \& Wrede}]{Bishop2003}
Bishop, S., Azuma, R.~E., Buchmann, L., {et~al.} 2003, \bibinfo{title}{$^{21}\mathrm{N}\mathrm{a}(p,\ensuremath{\gamma})^{22}\mathrm{M}\mathrm{g}$ Reaction and Oxygen-Neon Novae,} Phys. Rev. Lett., 90, 162501, \dodoi{10.1103/PhysRevLett.90.162501}

\bibitem[{J. {Casanova} {et~al.}(2010){Casanova}, {Jos{\'e}}, {Garc{\'\i}a-Berro}, {Calder}, \& {Shore}}]{Casanova2010}
{Casanova}, J., {Jos{\'e}}, J., {Garc{\'\i}a-Berro}, E., {Calder}, A., \& {Shore}, S.~N. 2010, \bibinfo{title}{{On mixing at the core-envelope interface during classical nova outbursts},} \aap, 513, L5, \dodoi{10.1051/0004-6361/201014178}

\bibitem[{J. {Casanova} {et~al.}(2011{\natexlab{a}}){Casanova}, {Jos{\'e}}, {Garc{\'\i}a-Berro}, {Calder}, \& {Shore}}]{Casanova2011a}
{Casanova}, J., {Jos{\'e}}, J., {Garc{\'\i}a-Berro}, E., {Calder}, A., \& {Shore}, S.~N. 2011{\natexlab{a}}, \bibinfo{title}{{Mixing in classical novae: a 2-D sensitivity study},} \aap, 527, A5, \dodoi{10.1051/0004-6361/201015895}

\bibitem[{J. {Casanova} {et~al.}(2016){Casanova}, {Jos{\'e}}, {Garc{\'\i}a-Berro}, \& {Shore}}]{Casanova2016}
{Casanova}, J., {Jos{\'e}}, J., {Garc{\'\i}a-Berro}, E., \& {Shore}, S.~N. 2016, \bibinfo{title}{{Three-dimensional simulations of turbulent convective mixing in ONe and CO classical nova explosions},} \aap, 595, A28, \dodoi{10.1051/0004-6361/201628707}

\bibitem[{J. {Casanova} {et~al.}(2011{\natexlab{b}}){Casanova}, {Jos{\'e}}, {Garc{\'\i}a-Berro}, {Shore}, \& {Calder}}]{Casanova2011b}
{Casanova}, J., {Jos{\'e}}, J., {Garc{\'\i}a-Berro}, E., {Shore}, S.~N., \& {Calder}, A.~C. 2011{\natexlab{b}}, \bibinfo{title}{{Kelvin-Helmholtz instabilities as the source of inhomogeneous mixing in nova explosions},} \nat, 478, 490, \dodoi{10.1038/nature10520}

\bibitem[{J. {Casanova} {et~al.}(2018){Casanova}, {Jos{\'e}}, \& {Shore}}]{Casanova2018}
{Casanova}, J., {Jos{\'e}}, J., \& {Shore}, S.~N. 2018, \bibinfo{title}{{Two-dimensional simulations of mixing in classical novae: The effect of white dwarf composition and mass},} \aap, 619, A121, \dodoi{10.1051/0004-6361/201833422}

\bibitem[{G. {Cescutti} {et~al.}(2018){Cescutti}, {Hirschi}, {Nishimura}, {Hartogh}, {Rauscher}, {Murphy}, \& {Cristallo}}]{Cescutti2018}
{Cescutti}, G., {Hirschi}, R., {Nishimura}, N., {et~al.} 2018, \bibinfo{title}{{Uncertainties in s-process nucleosynthesis in low-mass stars determined from Monte Carlo variations},} \mnras, 478, 4101, \dodoi{10.1093/mnras/sty1185}

\bibitem[{H.-L. {Chen} {et~al.}(2019){Chen}, {Woods}, {Yungelson}, {Piersanti}, {Gilfanov}, \& {Han}}]{Chen2019}
{Chen}, H.-L., {Woods}, T.~E., {Yungelson}, L.~R., {et~al.} 2019, \bibinfo{title}{{Comprehensive models of novae at metallicity Z = 0.02 and Z = 10$^{-4}$},} \mnras, 490, 1678, \dodoi{10.1093/mnras/stz2644}

\bibitem[{L. {Chomiuk} {et~al.}(2021){Chomiuk}, {Metzger}, \& {Shen}}]{Chomiuk2021}
{Chomiuk}, L., {Metzger}, B.~D., \& {Shen}, K.~J. 2021, \bibinfo{title}{{New Insights into Classical Novae},} \araa, 59, 391, \dodoi{10.1146/annurev-astro-112420-114502}

\bibitem[{F. Crameri(2023)Crameri}]{crameri}
Crameri, F. 2023, \bibinfo{title}{{Scientific colour maps},}, 8.0.1 Zenodo, \dodoi{10.5281/zenodo.8409685}

\bibitem[{P.~A. {Denissenkov} {et~al.}(2021){Denissenkov}, {Herwig}, {Perdikakis}, \& {Schatz}}]{Denissenkov2021}
{Denissenkov}, P.~A., {Herwig}, F., {Perdikakis}, G., \& {Schatz}, H. 2021, \bibinfo{title}{{The impact of (n,{\ensuremath{\gamma}}) reaction rate uncertainties of unstable isotopes on the i-process nucleosynthesis of the elements from Ba to W},} \mnras, 503, 3913, \dodoi{10.1093/mnras/stab772}

\bibitem[{T.~F. {Doyle} {et~al.}(2019){Doyle}, {Shara}, {Lessing}, \& {Zurek}}]{Doyle2019}
{Doyle}, T.~F., {Shara}, M.~M., {Lessing}, A.~M., \& {Zurek}, D. 2019, \bibinfo{title}{{A Hubble Space Telescope Survey for Novae in the Globular Clusters of M87},} \apj, 874, 65, \dodoi{10.3847/1538-4357/ab0490}

\bibitem[{A. Evans {et~al.}(2024)Evans, Banerjee, Geballe, Polin, Hsiao, Page, Woodward, \& Starrfield}]{Evans2024}
Evans, A., Banerjee, D. P.~K., Geballe, T.~R., {et~al.} 2024, \bibinfo{title}{Near-infrared spectroscopy of the LMC recurrent nova LMCN 1968-12a,} \mnras, \dodoi{10.1093/mnras/stae2711}

\bibitem[{J.~L. {Fisker} {et~al.}(2008){Fisker}, {Schatz}, \& {Thielemann}}]{Fisker2008}
{Fisker}, J.~L., {Schatz}, H., \& {Thielemann}, F.-K. 2008, \bibinfo{title}{{Explosive Hydrogen Burning during Type I X-Ray Bursts},} \apjs, 174, 261, \dodoi{10.1086/521104}

\bibitem[{S.~A. {Glasner} {et~al.}(1997){Glasner}, {Livne}, \& {Truran}}]{Glasner1997}
{Glasner}, S.~A., {Livne}, E., \& {Truran}, J.~W. 1997, \bibinfo{title}{{Reactive Flow in Nova Outbursts},} \apj, 475, 754, \dodoi{10.1086/303561}

\bibitem[{S.~A. {Glasner} {et~al.}(2007){Glasner}, {Livne}, \& {Truran}}]{Glasner2007}
{Glasner}, S.~A., {Livne}, E., \& {Truran}, J.~W. 2007, \bibinfo{title}{{Novae: The Evolution from Onset of Convection to the Runaway},} \apj, 665, 1321, \dodoi{10.1086/519234}

\bibitem[{S.~A. {Glasner} {et~al.}(2012){Glasner}, {Livne}, \& {Truran}}]{Glasner2012}
{Glasner}, S.~A., {Livne}, E., \& {Truran}, J.~W. 2012, \bibinfo{title}{{Convective overshoot mixing in Nova outbursts - the dependence on the composition of the underlyingwhitedwarf},} \mnras, 427, 2411, \dodoi{10.1111/j.1365-2966.2012.22103.x}

\bibitem[{S.~A. {Glasner} \& J.~W. {Truran}(2009){Glasner} \& {Truran}}]{Glassner2009}
{Glasner}, S.~A., \& {Truran}, J.~W. 2009, \bibinfo{title}{{Carbon-Nitrogen-Oxygen ``Breakout'' and Nucleosynthesis in Classical Novae},} \apjl, 692, L58, \dodoi{10.1088/0004-637X/692/1/L58}

\bibitem[{S. {Goriely} {et~al.}(2008){Goriely}, {Hilaire}, \& {Koning}}]{Goriely2008}
{Goriely}, S., {Hilaire}, S., \& {Koning}, A.~J. 2008, \bibinfo{title}{{Improved predictions of nuclear reaction rates with the TALYS reaction code for astrophysical applications},} \aap, 487, 767, \dodoi{10.1051/0004-6361:20078825}

\bibitem[{C.~R. Harris {et~al.}(2020)Harris, Millman, van~der Walt, Gommers, Virtanen, Cournapeau, Wieser, Taylor, Berg, Smith, Kern, Picus, Hoyer, van Kerkwijk, Brett, Haldane, del R{\'{i}}o, Wiebe, Peterson, G{\'{e}}rard-Marchant, Sheppard, Reddy, Weckesser, Abbasi, Gohlke, \& Oliphant}]{numpy}
Harris, C.~R., Millman, K.~J., van~der Walt, S.~J., {et~al.} 2020, \bibinfo{title}{Array programming with {NumPy},} \nat, 585, 357, \dodoi{10.1038/s41586-020-2649-2}

\bibitem[{T. {Hartwig} {et~al.}(2015){Hartwig}, {Bromm}, {Klessen}, \& {Glover}}]{Hartwig2015}
{Hartwig}, T., {Bromm}, V., {Klessen}, R.~S., \& {Glover}, S. C.~O. 2015, \bibinfo{title}{{Constraining the primordial initial mass function with stellar archaeology},} \mnras, 447, 3892, \dodoi{10.1093/mnras/stu2740}

\bibitem[{M. {Henze} {et~al.}(2013){Henze}, {Pietsch}, {Haberl}, {Della Valle}, {Riffeser}, {Sala}, {Hatzidimitriou}, {Hofmann}, {Hartmann}, {Koppenhoefer}, {Seitz}, {Williams}, {Hornoch}, {Itagaki}, {Kabashima}, {Nishiyama}, {Xing}, {Lee}, {Magnier}, \& {Chambers}}]{Henze2013}
{Henze}, M., {Pietsch}, W., {Haberl}, F., {et~al.} 2013, \bibinfo{title}{{Supersoft X-rays reveal a classical nova in the M 31 globular cluster Bol 126},} \aap, 549, A120, \dodoi{10.1051/0004-6361/201220196}

\bibitem[{V. {Hocd{\'e}} {et~al.}(2023){Hocd{\'e}}, {Smolec}, {Moskalik}, {Zi{\'o}{\l}kowska}, \& {Singh Rathour}}]{Hocde2023}
{Hocd{\'e}}, V., {Smolec}, R., {Moskalik}, P., {Zi{\'o}{\l}kowska}, O., \& {Singh Rathour}, R. 2023, \bibinfo{title}{{Metallicity estimations of MW, SMC, and LMC classical Cepheids from the shape of the V- and I-band light curves},} \aap, 671, A157, \dodoi{10.1051/0004-6361/202245038}

\bibitem[{J.~D. Hunter(2007)Hunter}]{matplotlib}
Hunter, J.~D. 2007, \bibinfo{title}{Matplotlib: A 2D graphics environment,} Computing in Science \& Engineering, 9, 90, \dodoi{10.1109/MCSE.2007.55}

\bibitem[{C. Iliadis(2015)Iliadis}]{Iliadis2015b}
Iliadis, C. 2015, Nuclear physics of stars (John Wiley \& Sons)

\bibitem[{C. {Iliadis} {et~al.}(2002){Iliadis}, {Champagne}, {Jos{\'e}}, {Starrfield}, \& {Tupper}}]{Iliadis2002}
{Iliadis}, C., {Champagne}, A., {Jos{\'e}}, J., {Starrfield}, S., \& {Tupper}, P. 2002, \bibinfo{title}{{The Effects of Thermonuclear Reaction-Rate Variations on Nova Nucleosynthesis: A Sensitivity Study},} \apjs, 142, 105, \dodoi{10.1086/341400}

\bibitem[{C. {Iliadis} \& A. {Coc}(2020){Iliadis} \& {Coc}}]{Iliadis2020}
{Iliadis}, C., \& {Coc}, A. 2020, \bibinfo{title}{{Thermonuclear Reaction Rates and Primordial Nucleosynthesis},} \apj, 901, 127, \dodoi{10.3847/1538-4357/abb1a3}

\bibitem[{C. {Iliadis} {et~al.}(2016){Iliadis}, {Karakas}, {Prantzos}, {Lattanzio}, \& {Doherty}}]{Iliadis2016}
{Iliadis}, C., {Karakas}, A.~I., {Prantzos}, N., {Lattanzio}, J.~C., \& {Doherty}, C.~L. 2016, \bibinfo{title}{{On Potassium and Other Abundance Anomalies of Red Giants in NGC 2419},} \apj, 818, 98, \dodoi{10.3847/0004-637X/818/1/98}

\bibitem[{C. {Iliadis} {et~al.}(2015){Iliadis}, {Longland}, {Coc}, {Timmes}, \& {Champagne}}]{Iliadis2015}
{Iliadis}, C., {Longland}, R., {Coc}, A., {Timmes}, F.~X., \& {Champagne}, A.~E. 2015, \bibinfo{title}{{Statistical methods for thermonuclear reaction rates and nucleosynthesis simulations},} Journal of Physics G Nuclear Physics, 42, 034007, \dodoi{10.1088/0954-3899/42/3/034007}

\bibitem[{J. {Jos{\'e}}(2016){Jos{\'e}}}]{Jose2016}
{Jos{\'e}}, J. 2016, {Stellar Explosions: Hydrodynamics and Nucleosynthesis} (CRC Press), \dodoi{10.1201/b19165}

\bibitem[{J. {Jos{\'e}} {et~al.}(2007){Jos{\'e}}, {Garc{\'\i}a-Berro}, {Hernanz}, \& {Gil-Pons}}]{Jose2007b}
{Jos{\'e}}, J., {Garc{\'\i}a-Berro}, E., {Hernanz}, M., \& {Gil-Pons}, P. 2007, \bibinfo{title}{{The First Nova Explosions},} \apjl, 662, L103, \dodoi{10.1086/519521}

\bibitem[{J. {Jos{\'e}} \& M. {Hernanz}(1998){Jos{\'e}} \& {Hernanz}}]{Jose1998}
{Jos{\'e}}, J., \& {Hernanz}, M. 1998, \bibinfo{title}{{Nucleosynthesis in Classical Novae: CO versus ONe White Dwarfs},} \apj, 494, 680, \dodoi{10.1086/305244}

\bibitem[{J. {Jos{\'e}} \& S.~N. {Shore}(2008){Jos{\'e}} \& {Shore}}]{Jose2008}
{Jos{\'e}}, J., \& {Shore}, S.~N. 2008, in Classical Novae, ed. M.~F. {Bode} \& A.~{Evans}, Vol.~43 (Cambridge University Press), 121--149, \dodoi{10.1017/CBO9780511536168.008}

\bibitem[{J. {Jos{\'e}} {et~al.}(2020){Jos{\'e}}, {Shore}, \& {Casanova}}]{Jose2020}
{Jos{\'e}}, J., {Shore}, S.~N., \& {Casanova}, J. 2020, \bibinfo{title}{{123-321 models of classical novae},} \aap, 634, A5, \dodoi{10.1051/0004-6361/201936893}

\bibitem[{S.~C. {Keller} {et~al.}(2014){Keller}, {Bessell}, {Frebel}, {Casey}, {Asplund}, {Jacobson}, {Lind}, {Norris}, {Yong}, {Heger}, {Magic}, {da Costa}, {Schmidt}, \& {Tisserand}}]{Keller2014}
{Keller}, S.~C., {Bessell}, M.~S., {Frebel}, A., {et~al.} 2014, \bibinfo{title}{{A single low-energy, iron-poor supernova as the source of metals in the star SMSS J031300.36-670839.3},} \nat, 506, 463, \dodoi{10.1038/nature12990}

\bibitem[{A.~J. {Kemp} {et~al.}(2024){Kemp}, {Karakas}, {Casey}, {C{\^o}t{\'e}}, {Izzard}, \& {Osborn}}]{Kemp2024}
{Kemp}, A.~J., {Karakas}, A.~I., {Casey}, A.~R., {et~al.} 2024, \bibinfo{title}{{Nova contributions to the chemical evolution of the Milky Way},} \aap, 689, A222, \dodoi{10.1051/0004-6361/202450800}

\bibitem[{R.~S. {Klessen} \& S.~C.~O. {Glover}(2023){Klessen} \& {Glover}}]{Klessen2023}
{Klessen}, R.~S., \& {Glover}, S. C.~O. 2023, \bibinfo{title}{{The First Stars: Formation, Properties, and Impact},} \araa, 61, 65, \dodoi{10.1146/annurev-astro-071221-053453}

\bibitem[{T. Kluyver {et~al.}(2016)Kluyver, Ragan-Kelley, P{\'e}rez, Granger, Bussonnier, Frederic, Kelley, Hamrick, Grout, Corlay, Ivanov, Avila, Abdalla, \& Willing}]{jupyter}
Kluyver, T., Ragan-Kelley, B., P{\'e}rez, F., {et~al.} 2016, in Positioning and Power in Academic Publishing: Players, Agents and Agendas, ed. F.~Loizides \& B.~Schmidt, IOS Press, 87 -- 90

\bibitem[{K. Lodders(2020)Lodders}]{Lodders2020}
Lodders, K. 2020, \bibinfo{title}{Solar Elemental Abundances,} Oxford Research Encyclopedia of Planetary Science, \dodoi{10.1093/acrefore/9780190647926.013.145}

\bibitem[{R. Longland(2012)Longland}]{Longland2012}
Longland, R. 2012, \bibinfo{title}{Recommendations for Monte Carlo nucleosynthesis sampling,} \aap, 548, A30, \dodoi{10.1051/0004-6361/201220386}

\bibitem[{R. Longland {et~al.}(2010)Longland, Iliadis, Champagne, Newton, Ugalde, Coc, \& Fitzgerald}]{Longland2010}
Longland, R., Iliadis, C., Champagne, A., {et~al.} 2010, \bibinfo{title}{Charged-particle thermonuclear reaction rates: I. Monte Carlo method and statistical distributions,} \nphysa, 841, 1, \dodoi{10.1016/j.nuclphysa.2010.04.008}

\bibitem[{R. {Longland} {et~al.}(2014){Longland}, {Martin}, \& {Jos{\'e}}}]{Longland2014}
{Longland}, R., {Martin}, D., \& {Jos{\'e}}, J. 2014, \bibinfo{title}{{Performance improvements for nuclear reaction network integration},} \aap, 563, A67, \dodoi{10.1051/0004-6361/201321958}

\bibitem[{R. {Maiolino} {et~al.}(2024){Maiolino}, {{\"U}bler}, {Perna}, {Scholtz}, {D'Eugenio}, {Witten}, {Laporte}, {Witstok}, {Carniani}, {Tacchella}, {Baker}, {Arribas}, {Nakajima}, {Eisenstein}, {Bunker}, {Charlot}, {Cresci}, {Curti}, {Curtis-Lake}, {de Graaff}, {Egami}, {Ji}, {Johnson}, {Kumari}, {Looser}, {Maseda}, {Nelson}, {Robertson}, {Rodr{\'\i}guez Del Pino}, {Sandles}, {Simmonds}, {Smit}, {Sun}, {Venturi}, {Williams}, \& {Willmer}}]{Maiolino2024}
{Maiolino}, R., {{\"U}bler}, H., {Perna}, M., {et~al.} 2024, \bibinfo{title}{{JADES. Possible Population III signatures at z = 10.6 in the halo of GN-z11},} \aap, 687, A67, \dodoi{10.1051/0004-6361/202347087}

\bibitem[{M.~R. {Mumpower} {et~al.}(2016){Mumpower}, {Surman}, {McLaughlin}, \& {Aprahamian}}]{Mumpower2016}
{Mumpower}, M.~R., {Surman}, R., {McLaughlin}, G.~C., \& {Aprahamian}, A. 2016, \bibinfo{title}{{The impact of individual nuclear properties on r-process nucleosynthesis},} Progress in Particle and Nuclear Physics, 86, 86, \dodoi{10.1016/j.ppnp.2015.09.001}

\bibitem[{K. {Nakajima} \& R. {Maiolino}(2022){Nakajima} \& {Maiolino}}]{Nakajima2022}
{Nakajima}, K., \& {Maiolino}, R. 2022, \bibinfo{title}{{Diagnostics for PopIII galaxies and direct collapse black holes in the early universe},} \mnras, 513, 5134, \dodoi{10.1093/mnras/stac1242}

\bibitem[{N. {Nishimura} {et~al.}(2019){Nishimura}, {Rauscher}, {Hirschi}, {Cescutti}, {Murphy}, \& {Fr{\"o}hlich}}]{Nishimura2019}
{Nishimura}, N., {Rauscher}, T., {Hirschi}, R., {et~al.} 2019, \bibinfo{title}{{Uncertainties in {\ensuremath{\nu}}p-process nucleosynthesis from Monte Carlo variation of reaction rates},} \mnras, 489, 1379, \dodoi{10.1093/mnras/stz2104}

\bibitem[{A. Parikh {et~al.}(2008)Parikh, Jos\'e, Moreno, \& Iliadis}]{Parikh2008}
Parikh, A., Jos\'e, J., Moreno, F., \& Iliadis, C. 2008, \bibinfo{title}{The effects of variations in nuclear processes on type I X-ray burst nucleosynthesis,} \apjs, 178, 110, \dodoi{10.1086/589879}

\bibitem[{F. Pedregosa {et~al.}(2011)Pedregosa, Varoquaux, Gramfort, Michel, Thirion, Grisel, Blondel, Prettenhofer, Weiss, Dubourg, Vanderplas, Passos, Cournapeau, Brucher, Perrot, \& Duchesnay}]{scikit-learn}
Pedregosa, F., Varoquaux, G., Gramfort, A., {et~al.} 2011, \bibinfo{title}{Scikit-learn: Machine Learning in {P}ython,} Journal of Machine Learning Research, 12, 2825

\bibitem[{F. P\'erez \& B.~E. Granger(2007)P\'erez \& Granger}]{ipython}
P\'erez, F., \& Granger, B.~E. 2007, \bibinfo{title}{{IP}ython: a System for Interactive Scientific Computing,} Computing in Science and Engineering, 9, 21, \dodoi{10.1109/MCSE.2007.53}

\bibitem[{L. {Piersanti} {et~al.}(2000){Piersanti}, {Cassisi}, {Iben}, \& {Tornamb{\'e}}}]{Piersanti2000}
{Piersanti}, L., {Cassisi}, S., {Iben}, Jr., I., \& {Tornamb{\'e}}, A. 2000, \bibinfo{title}{{Hydrogen-Accreting Carbon-Oxygen White Dwarfs of Low Mass: Thermal and Chemical Behavior of Burning Shells},} \apj, 535, 932, \dodoi{10.1086/308885}

\bibitem[{A. {Psaltis} {et~al.}(2022){Psaltis}, {Arcones}, {Montes}, {Mohr}, {Hansen}, {Jacobi}, \& {Schatz}}]{Psaltis2022}
{Psaltis}, A., {Arcones}, A., {Montes}, F., {et~al.} 2022, \bibinfo{title}{{Constraining Nucleosynthesis in Neutrino-driven Winds: Observations, Simulations, and Nuclear Physics},} \apj, 935, 27, \dodoi{10.3847/1538-4357/ac7da7}

\bibitem[{T. {Rauscher} {et~al.}(2016){Rauscher}, {Nishimura}, {Hirschi}, {Cescutti}, {Murphy}, \& {Heger}}]{Rauscher2016}
{Rauscher}, T., {Nishimura}, N., {Hirschi}, R., {et~al.} 2016, \bibinfo{title}{{Uncertainties in the production of p nuclei in massive stars obtained from Monte Carlo variations},} \mnras, 463, 4153, \dodoi{10.1093/mnras/stw2266}

\bibitem[{C. {Ritossa} {et~al.}(1996){Ritossa}, {Garcia-Berro}, \& {Iben}}]{Ritossa1996}
{Ritossa}, C., {Garcia-Berro}, E., \& {Iben}, Icko, J. 1996, \bibinfo{title}{{On the Evolution of Stars That Form Electron-degenerate Cores Processed by Carbon Burning. II. Isotope Abundances and Thermal Pulses in a 10 M$_{sun}$ Model with an ONe Core and Applications to Long-Period Variables, Classical Novae, and Accretion-induced Collapse},} \apj, 460, 489, \dodoi{10.1086/176987}

\bibitem[{A.~L. {Sallaska} {et~al.}(2013){Sallaska}, {Iliadis}, {Champange}, {Goriely}, {Starrfield}, \& {Timmes}}]{Sallaska2013}
{Sallaska}, A.~L., {Iliadis}, C., {Champange}, A.~E., {et~al.} 2013, \bibinfo{title}{{STARLIB: A Next-generation Reaction-rate Library for Nuclear Astrophysics},} \apjs, 207, 18, \dodoi{10.1088/0067-0049/207/1/18}

\bibitem[{H.~B. {Sawyer}(1938){Sawyer}}]{Sawyer1938}
{Sawyer}, H.~B. 1938, \bibinfo{title}{{The Bright Nova of 1860 in the Globular Cluster Messier 80. and its Relation to Supernovae},} \jrasc, 32, 69

\bibitem[{H. {Sawyer Hogg} \& A. {Wehlau}(1964){Sawyer Hogg} \& {Wehlau}}]{Sawyer1964}
{Sawyer Hogg}, H., \& {Wehlau}, A. 1964, \bibinfo{title}{{A Photographic Nova in the Globular Cluster Messier 14},} \jrasc, 58, 163

\bibitem[{H. {Schatz} {et~al.}(1998){Schatz}, {Aprahamian}, {Goerres}, {Wiescher}, {Rauscher}, {Rembges}, {Thielemann}, {Pfeiffer}, {Moeller}, {Kratz}, {Herndl}, {Brown}, \& {Rebel}}]{Schatz1998}
{Schatz}, H., {Aprahamian}, A., {Goerres}, J., {et~al.} 1998, \bibinfo{title}{{rp-Process Nucleosynthesis at Extreme Temperature and Density Conditions},} \physrep, 294, 167, \dodoi{10.1016/S0370-1573(97)00048-3}

\bibitem[{A.~W. {Shafter}(2017){Shafter}}]{Shafter2017}
{Shafter}, A.~W. 2017, \bibinfo{title}{{The Galactic Nova Rate Revisited},} \apj, 834, 196, \dodoi{10.3847/1538-4357/834/2/196}

\bibitem[{K.~J. {Shen} \& L. {Bildsten}(2007){Shen} \& {Bildsten}}]{Shen2007}
{Shen}, K.~J., \& {Bildsten}, L. 2007, \bibinfo{title}{{Thermally Stable Nuclear Burning on Accreting White Dwarfs},} \apj, 660, 1444, \dodoi{10.1086/513457}

\bibitem[{K.~J. {Shen} \& L. {Bildsten}(2009){Shen} \& {Bildsten}}]{Shen2009}
{Shen}, K.~J., \& {Bildsten}, L. 2009, \bibinfo{title}{{The Effect of Composition on Nova Ignitions},} \apj, 692, 324, \dodoi{10.1088/0004-637X/692/1/324}

\bibitem[{E. {Spitoni} {et~al.}(2021){Spitoni}, {Verma}, {Silva Aguirre}, {Vincenzo}, {Matteucci}, {Vai{\v{c}}ekauskait{\.{e}}}, {Palla}, {Grisoni}, \& {Calura}}]{Spitoni2021}
{Spitoni}, E., {Verma}, K., {Silva Aguirre}, V., {et~al.} 2021, \bibinfo{title}{{APOGEE DR16: A multi-zone chemical evolution model for the Galactic disc based on MCMC methods},} \aap, 647, A73, \dodoi{10.1051/0004-6361/202039864}

\bibitem[{A. {Stacy} \& V. {Bromm}(2013){Stacy} \& {Bromm}}]{Stacy2013}
{Stacy}, A., \& {Bromm}, V. 2013, \bibinfo{title}{{Constraining the statistics of Population III binaries},} \mnras, 433, 1094, \dodoi{10.1093/mnras/stt789}

\bibitem[{A. {Stacy} {et~al.}(2016){Stacy}, {Bromm}, \& {Lee}}]{Stacy2016}
{Stacy}, A., {Bromm}, V., \& {Lee}, A.~T. 2016, \bibinfo{title}{{Building up the Population III initial mass function from cosmological initial conditions},} \mnras, 462, 1307, \dodoi{10.1093/mnras/stw1728}

\bibitem[{S. {Starrfield} {et~al.}(2020){Starrfield}, {Bose}, {Iliadis}, {Hix}, {Woodward}, \& {Wagner}}]{Starrfield2020}
{Starrfield}, S., {Bose}, M., {Iliadis}, C., {et~al.} 2020, \bibinfo{title}{{Carbon-Oxygen Classical Novae Are Galactic $^{7}$Li Producers as well as Potential Supernova Ia Progenitors},} \apj, 895, 70, \dodoi{10.3847/1538-4357/ab8d23}

\bibitem[{S. {Starrfield} {et~al.}(2008){Starrfield}, {Iliadis}, \& {Hix}}]{Starrfield2008}
{Starrfield}, S., {Iliadis}, C., \& {Hix}, W.~R. 2008, in Classical Novae, ed. M.~F. {Bode} \& A.~{Evans}, Vol.~43 (Cambridge University Press), 77--101, \dodoi{10.1017/CBO9780511536168.006}

\bibitem[{S. {Starrfield} {et~al.}(2016){Starrfield}, {Iliadis}, \& {Hix}}]{Starrfield2016}
{Starrfield}, S., {Iliadis}, C., \& {Hix}, W.~R. 2016, \bibinfo{title}{{The Thermonuclear Runaway and the Classical Nova Outburst},} \pasp, 128, 051001, \dodoi{10.1088/1538-3873/128/963/051001}

\bibitem[{S. {Starrfield} {et~al.}(2000){Starrfield}, {Schwarz}, {Truran}, \& {Sparks}}]{Starrfield2000}
{Starrfield}, S., {Schwarz}, G., {Truran}, J.~W., \& {Sparks}, W.~M. 2000, in American Institute of Physics Conference Series, Vol. 522, Cosmic Explosions: Tenth AstroPhysics Conference, ed. S.~S. {Holt} \& W.~W. {Zhang} (AIP), 379--382, \dodoi{10.1063/1.1291739}

\bibitem[{S. {Starrfield} {et~al.}(2012){Starrfield}, {Timmes}, {Iliadis}, {Hix}, {Arnett}, {Meakin}, \& {Sparks}}]{Starrfield2012}
{Starrfield}, S., {Timmes}, F.~X., {Iliadis}, C., {et~al.} 2012, \bibinfo{title}{{Hydrodynamic Studies of the Evolution of Recurrent, Symbiotic and Dwarf Novae: the White Dwarf Components are Growing in Mass},} Baltic Astronomy, 21, 76, \dodoi{10.1515/astro-2017-0361}

\bibitem[{L. {van Wormer} {et~al.}(1994){van Wormer}, {G{\"o}rres}, {Iliadis}, {Wiescher}, \& {Thielemann}}]{vanWormer1994}
{van Wormer}, L., {G{\"o}rres}, J., {Iliadis}, C., {Wiescher}, M., \& {Thielemann}, F.~K. 1994, \bibinfo{title}{{Reaction Rates and Reaction Sequences in the rp-Process},} \apj, 432, 326, \dodoi{10.1086/174572}

\bibitem[{R.~V. {Wagoner}(1969){Wagoner}}]{Wagoner1969}
{Wagoner}, R.~V. 1969, \bibinfo{title}{{Synthesis of the Elements Within Objects Exploding from Very High Temperatures},} \apjs, 18, 247, \dodoi{10.1086/190191}

\bibitem[{R.~K. {Wallace} \& S.~E. {Woosley}(1981){Wallace} \& {Woosley}}]{Wallace1981}
{Wallace}, R.~K., \& {Woosley}, S.~E. 1981, \bibinfo{title}{{Explosive hydrogen burning},} \apjs, 45, 389, \dodoi{10.1086/190717}

\bibitem[{M. {Wiescher} {et~al.}(2010){Wiescher}, {G{\"o}rres}, {Uberseder}, {Imbriani}, \& {Pignatari}}]{Wiescher2010}
{Wiescher}, M., {G{\"o}rres}, J., {Uberseder}, E., {Imbriani}, G., \& {Pignatari}, M. 2010, \bibinfo{title}{{The Cold and Hot CNO Cycles},} Annual Review of Nuclear and Particle Science, 60, 381, \dodoi{10.1146/annurev.nucl.012809.104505}

\bibitem[{E. Zinner(2014)Zinner}]{Zinner2014}
Zinner, E. 2014, \bibinfo{title}{Presolar grains,} Meteorites and cosmochemical processes, 1, 181, \dodoi{10.1016/B978-0-08-095975-7.00101-7}

\end{thebibliography}
\bibliographystyle{aasjournal}



\end{document}